\documentclass[twocolumn,superscriptaddress,showpacs]{revtex4}

\usepackage{mathrsfs}
\usepackage[centertags]{amsmath}
\usepackage{amssymb}
\usepackage{amsthm}
\usepackage{amsmath}
\usepackage{CJK}
\usepackage{graphicx,subfigure,epstopdf}
\usepackage{indentfirst}
\usepackage{fancyhdr}
\usepackage{bm} 
\usepackage{esint}
\newcommand{\ie}{\emph{i.e.}}

\newcommand{\ER}{Erd\H{o}s-R\'{e}nyi}

\begin{document}

\title{Explosive percolation with multiple giant components}
\author{Wei Chen}
\email{chwei@ucdavis.edu}
\affiliation{School of Mathematical Sciences, Peking University, Beijing, China}
\affiliation{University of California, Davis, CA 95616, USA}
\author{Raissa M. D'Souza}
\email{raissa@cse.ucdavis.edu}
\affiliation{University of California, Davis, CA 95616, USA}
\affiliation{Santa Fe Institute, 1399 Hyde Park Road, Santa Fe, New Mexico 87501, USA}

\begin{abstract}
We generalize the random graph evolution process of Bohman, Frieze, and Wormald [T. Bohman, A. Frieze, and N. C. Wormald, Random Struct. Algorithms, {\bf 25}, 432 (2004)].  
Potential edges, sampled uniformly at random from the complete graph, 
are considered one at a time and either added to the graph or rejected provided that the fraction of accepted edges is never smaller than a decreasing function asymptotically approaching the value $\alpha = 1/2$. We show that multiple giant components appear simultaneously in  a strongly discontinuous percolation transition and 
remain distinct. 
Furthermore, tuning the value of $\alpha$ determines the number of 
such components 
with smaller $\alpha$ leading to an increasingly delayed and more explosive transition.  
The location of the critical point and strongly discontinuous nature are not affected if
only edges which span components are sampled.
%
\end{abstract}


\maketitle

The percolation phase transition models 
the onset of large-scale connectivity in lattices or networks,  in systems ranging from porous media, to resistor networks, to epidemic spreading~\cite{StaufferPercBook,Sahimi,satorrasPRL,moorePRE}.
Percolation was considered a robust second-order transition until a variant with a choice between edges was shown to result in a seemingly discontinuous transition~\cite{EPScience}.
Subsequent studies have shown similar results for scale-free networks~\cite{ChoPRL09,RadFortPRL09}, lattices~\cite{ziffPRL09,Rziffnew}, local cluster aggregation models~\cite{RaissaMichael}, single-edge addition models~\cite{ChoKahng,Manna}, and models which control only the largest component~\cite{Araujo}.  It seems a fundamental requirement that in the sub-critical regime the evolution mechanism produces many clusters which are relatively large, though sublinear, in size~\cite{powderkeg09,RaissaMichael,ChoKahng}. 
Most recently, the notions of ``strongly" versus ``weakly" discontinuous transitions have been introduced~\cite{Timme2010}, with the model studied in~\cite{EPScience} showing weakly discontinuous characteristics, while an idealized deterministic ``most explosive'' model~\cite{powderkeg09,Makse2009} is  
strongly discontinuous. 
Here we analyze and extend a related model by Bohman, Frieze and Wormald (BFW)~\cite{T. Bohman and  A. Frieze}, which predates the more recent work, and show that surprisingly~\cite{SpencerAMS2010}, multiple stable giant components can coexist and that the percolation transition is strongly discontinuous.  

The ``most explosive" deterministic process~\cite{powderkeg09,Makse2009,Timme2010}
begins with $n$ isolated nodes, with $n$ set to a power of two for convenience. Edges that connect pairs of isolated nodes are added sequentially, creating components of size $k=2$,  until no isolated nodes remain. The cutoff $k$ is then doubled and edges leading to components of size $k=4$ are added sequentially, until all components have size $k=4$.  $k$ is then doubled yet again and the process iterated. By the end of the phase $k=n/2$, only two components remain, each with size $n/2$.  The addition of the next edge connects those two components during which the size of the largest component jumps in value by $n/2$.  Letting $t$ denote the number of edges added to the graph, we define the critical $t_c$ as the single edge who's addition produces the largest jump in size of the largest component denoted $\Delta C_{max}$; here $t_{c} = (n-1)$, with $\Delta C_{max} = n/2$. 

The BFW model begins with a collection of $n$ isolated nodes (with $n$ any integer) and also proceeds in phases starting with $k=2$. Edges are sampled one-at-a-time, uniformly at random from the complete graph. If an edge would lead to formation of a component of size less than or equal to $k$ it is accepted.  
Otherwise the edge is rejected provided that 
the fraction of accepted edges is greater than or equal to a function $g(k)=1/2+(2k)^{-1/2}$.  
If the accepted fraction is not sufficiently large, the phase is augmented to $k+1$ repeatedly until either the edge can be accommodated or $g(k)$ decreases sufficiently that the edge can be rejected.  
Explicit details are given below.   Asymptotically, $\lim_{k\rightarrow\infty} g(k)=\alpha$ with $\alpha =1/2$.

BFW established rigorous results whereby setting $g(200)=1/2$, all components are no larger than $k=200$ nodes (\ie, no giant component exists) when $m=0.96689n$ edges out of $2m$ sequentially sampled random edges have been added to graph.
They further establish that
a giant component must exist by the time $m=c^*n$ out of $2m$  sampled edges have been added, 
with $c^{*} \in [0.9792,0.9793]$.  
Yet, they did not analyze the details of the percolation transition. We show that their model leads to the simultaneous emergence of two giant components (each of size greater than 40\% of all the nodes), and show analytically the stability of the two giants throughout the subsequent graph evolution. We then generalize the BFW model by allowing the asymptotic fraction of accepted edges, $\alpha$,  to be a parameter and show that $\alpha$ determines the number of stable giant components that emerge and that, in general, smaller values of $\alpha$ lead to a more delayed and more explosive transition.

Stating the BFW algorithm in detail, let $k$ denote the stage and $n$ the number of nodes. Let $u$ denote the total number of edges sampled, $A$ the set of accepted edges (initialized to $A =\emptyset$), and $t=|A|$ the number of accepted edges. At each step $u$, an edge $e_{u}$ is sampled uniformly at random from the complete graph generated by the $n$ nodes, and the following algorithm iterated:  

{\tt Set $l=$ maximum size component in $A \cup \{e_{u}\}$

\ \ \ if $\left(l\leq k\right) \{$
 
\ \ \ \ \ \ $A \leftarrow A \cup \{e_{u}\}$
 
 \ \ \ \ \ \ $u\leftarrow u+1 \ \}$
  
\ \ \ else if $\left(t  / u < g(k)\right) \{$
 $k\leftarrow k+1\ \}$
 
\ \ \  else $\{\ u \leftarrow u+1 \ \}$

}

\vspace{1mm}
\noindent
Thus while $t/u < g(k)$, $k$ is augmented repeatedly until either  $k$ becomes large enough that edge $e_u$ is accepted or $g(k)$ decreases sufficiently that edge $e_u$ can be rejected at which point step $u$ ends.  Note $g(k)=1$ requires that all edges be accepted, equivalent to \ER~\cite{ER}. 
 
\begin{figure}
 \includegraphics[width=0.5\textwidth]{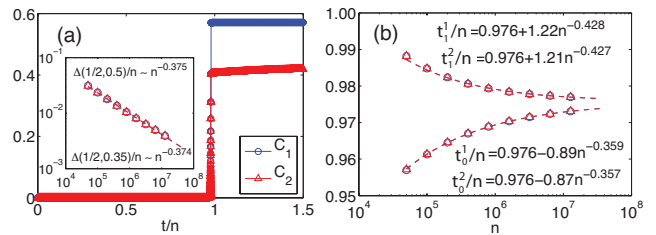}
 \caption{
 The BFW model. (a) $C_{1}$ and $C_{2}$ versus edge density, $t/n$, for $n=10^{6}$, showing the emergence of two stable giant components. 
Inset: $\Delta(1/2,0.5)/n, \Delta(1/2,0.35)/n$ for $C_{1}$ and $C_{2}$ versus $n$.  
(b) The lower and upper boundaries of $\Delta/n$ for both $C_1$ and $C_2$,  
where data points are averages over 200 to 2000 realizations and dashed lines are best fits.
}\label{fig:as}
  \end{figure}

We numerically implement the BFW model, and measure the fraction of nodes in both the largest and second largest component, denoted $C_{1}$ and $C_{2}$, as a function of edge density $t/n$. As shown in Fig.~\ref{fig:as}(a), two giants appear at the same critical point and remain distinct. 
To establish that the BFW model shows a seemingly discontinuous transition we apply the numerical method introduced in~\cite{EPScience}. 
Let $t^{i}_{0}(n)$ denote the last accepted edge for which $C_{i}n\leq{n^{\gamma}}$ and $t^{i}_{1}(n)$ the first accepted edge with $C_{i}n\geq An$, where $n$ is system size and $\gamma$ and $A$ are parameters. $\Delta^i(\gamma,A)=t^{i}_{1}(n)- t^{i}_{0}(n)$ denotes the number of accepted edges required between these two points.  As shown inset to Fig.~\ref{fig:as}(a), $\Delta^i(\gamma,A)/n$ is sublinear in $n$
and $t^{i}_{0}(n)/n$ and $t^{i}_{1}(n)/n$ converge to same limiting value $t_c/n=0.976$ (Fig.~\ref{fig:as}(b)) for both $C_1$ and $C_2$.

\begin{figure}
 
\includegraphics[width=0.50\textwidth]{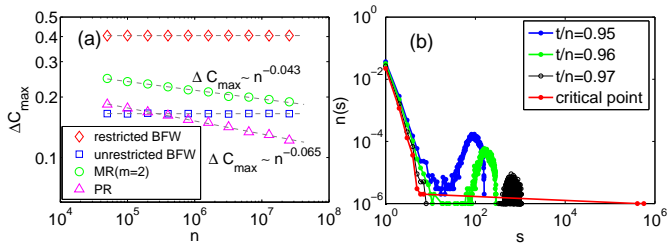}
 
\caption{
(a) $\Delta C_{\rm max}$, the biggest single edge increase in $C_{1}$, is independent of system size $n$ for BFW and for the restricted BFW process, indicating these transitions are strongly discontinuous. 
(b) Evolution of the distribution of $n(s)$, the fraction of components of size $s$, for the BFW model.}
 \label{fig:bs}
 \end{figure}

The discontinuous nature is made more explicit in Fig.~\ref{fig:bs}(a) showing $\Delta C_{\rm max}$, the largest increase of the largest component due to addition of a single edge, versus $n$ (blue squares are BFW).  
$\Delta C_{\rm max}$ is independent of $n$ (strongly discontinuous). Essentially the same value of $\Delta C_{\rm max}$ is always observed and results from the second and third components merging together to overtake what was previously the largest.  The model studied in~\cite{EPScience} (PR) shows a decrease with $n$ (weakly discontinuous), with scaling  $n^{-0.065}$ as also recently observed in~\cite{Manna,Timme2010}. 
 
The key to coexisting multiple giant components is the high probability of sampling internal-cluster links in the super-critical region which, by definition, do not increase the component size. We formalize this by first
introducing a function $P(k,t,n)$ defined as the probability of sampling a random link which leads to a component of size no larger than stage $k$ at step $t$ for system size $n$:
\begin{equation}
P(k,t,n)=\sum_{i}C_{i}^{2}+2\sum_{C_{i}+C_{j}\leq k/n}C_{i}C_{j}
\label{eqn:Pktn}
\end{equation}
where $C_{i}$ denotes the fraction of nodes in component $i$. The first term on the right-hand side is the probability of randomly sampling internal-cluster links in all components. The second term is the probability of sampling spanning-cluster links which lead to a component of size no larger than $k$.  This is valid for any configuration in phase $k$.  
We also consider $S(k,n)$, the the probability of sampling random links which lead to components of size no larger than $k$ if all possible spanning-cluster links are added in stage $k$. 
Thus
\begin{equation}
S(k,n)=\sum_{i}C_{i}^{2}.
\label{eqn:Skn}
\end{equation}
(Note the values of the $C_i$'s in Eq.~\ref{eqn:Pktn} and~\ref{eqn:Skn} can differ from each other.)
For any specific stage $k$,  it is easy to show that  
$P(k,t,n)\geq S(k,n)$
since, if $t$ increases, $P(k,t,n)$ can only decrease or stay the same. More explicitly, if an internal-cluster link is added then $P(k,t,n)$ is invariant, while if a spanning-cluster link is added between components $i$ and $j$ then the first term increases by $(C_{i}+C_{j})^{2}-(C_{i}^{2}+C_{j}^{2})=2C_{i}C_{j}$ and second term decreases by at least $2C_{i}C_{j}$. (Additional decreases result if there exist components $l$ 
satisfying $C_{i}+C_{l}\leq k/n$, but with $(C_{i}+C_{j})+C_{l}> k/n$).  

Focusing now on the critical region, let $k^{*}$ denote the value of $k$ at $t_c$. 
Numerical results for a variety of system sizes show that at $t_c$ when $n>10^{6}$, 
$k^{*}/n\sim 0.570$, 
$C_{1}\sim 0.570$ and $C_{2}\sim 0.405$ with error bars of order $\mathcal{O}(10^{-4})$ 
obtained over $30$ to $300$ realizations dependent on $n$. Thus the remaining components have total size density $\sum_{i\geq 3}C_{i}=0.025$.
We can establish a uniform lower bound on  $S(k,n)$ for all $k\ge k^*$ using the simple intuition that under the normalization condition 
$\sum_{i}C_{i}=1$, $S(k,n)=\sum_{i}C_{i}^{2}$ 
is minimized when the number of components are as numerous as possible and of similar size. 
Given that $\sum_{i\ge 3} C_{i} < C_{1}-C_{2}$, the 
lower bound on $S(k,n)$ is if all small components connect to $C_2$.  Then $P(k,t,n)\geq S(k,n)\geq C_{1}^{2}+(C_{2}+\sum_{i\geq 3}C_{i})^{2} =0.570^{2}+(0.405+0.025)^{2}\sim0.510$, so for any stage 
$k \geq k^{*}(n)$, we have 
\begin{equation}
P(k,t,n)>\alpha=1/2.
\label{eqn:pgalpha}
\end{equation}
So for $k \ge k^*$, the expected fraction of accepted links approaches a positive value strictly larger than $\alpha$. 

Having established that in expectation $P(k,t,n) > \alpha$, for $k\ge k^*$, we need to explicitly consider what happens if an edge connecting the two giant components is sampled in this regime.
Here $(C_{1}+C_{2})>k/n\geq k^{*}/n$ and, by definition, $t/u \ge g(k)$. Consider the case when edge $e_{u+1}$ connects $C_1$ and $C_2$.  If $t/(u+1) \ge g(k)$ the edge is simply rejected.  But if $t/(u+1) < g(k)$ either $k$ needs to increase until the edge is accommodated, or (as we show next) a small increase in $k$ quickly leads to $t/(u+1) \ge g(k)$.  Setting $t/u$ to the smallest value possible:  
\begin{equation}
\frac{t}{u}=\frac{1}{2}+\sqrt{\frac{1}{2k}}
\label{eqn:tu}
\end{equation}
Differentiating both sides in Eq.~(\ref{eqn:tu}) by $k$ we find that
\begin{equation}
\frac{du}{d(k/n)}=\frac{1}{2\sqrt{2}(1/2+\sqrt{1/2k})^{2}}\frac{nt}{k^{3/2}}.
\label{eqn:dudk}
\end{equation} 
After the critical point we know that 
$t \sim \mathcal{O}(n)$ and the stage $k\sim C_{1}n\sim \mathcal{O}(n)$. Thus from Eq.~(\ref{eqn:dudk}) it follows that $\frac{du}{d(k/n)}\sim \mathcal{O}(n^{1/2})$  
as $n\rightarrow \infty$  
implying that an $\mathcal{O}(n^{-1/2})$ increase in $k/n$  results in $t/(u+1)>g(k)$, so the link which would lead to merging $C_1$ and $C_2$ is rejected and the two giant components are stable throughout the subsequent evolution. 
We verify this numerically. 
Letting $\bar{k}(n)$ denote the largest value of the stage ever attained for system size $n$, we find
$(\bar{k}(n)-k^*(n))/n \sim n^{-\gamma}$ with $\gamma=0.46\pm{0.03}$, and as $n \rightarrow \infty$,  $k^{*}(n)/n$ and $\bar{k}(n)/n$ converge to the same limiting value of approximately 0.570.

The BFW model samples edges uniformly at random from the complete graph.  If we restrict the process to sampling only edges that span distinct clusters, we observe that two components with the same $C_1=0.570$ and $C_2=0.405$ values coexist for several edge additions before merging together. When they do merge the largest jump in $C_1$, equal to the size of $C_2$, 
occurs.  This is a strongly discontinuous transition 
as shown in Fig.~\ref{fig:bs}(a) (the red diamonds) with jump size equal to 0.405.

\begin{figure}
\includegraphics[width=0.50\textwidth]{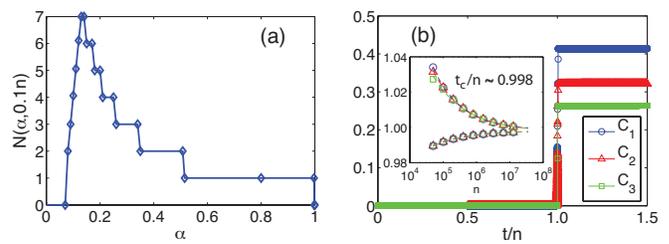} 
\caption{
(a) Number of stable giant components of size greater than $0.1n$ versus $\alpha$ for $n=10^{6}$. 
(b) When $\alpha=0.3$, three giant components emerge simultaneously. Inset shows 
convergence of the upper and lower boundaries of $\Delta/n$ for $C_{1},C_{2},C_{3}$, analogous to Fig.~\ref{fig:as}(b).
}
\label{fig:cs}
\end{figure}

We now generalize the BFW model so that $g(k)=\alpha+(2k)^{-1/2}$ 
(\ie, the asymptotic fraction of accepted links is now a parameter $\alpha$). 
For the unrestricted process (sampling from the complete graph) we find that $\alpha$ controls the number of stable giant components.  
Let $N(\alpha,m)$ denote the number of stable giant components with size larger than $m$ which appear at the critical point and remain throughout the subsequent evolution. 
Fig.~\ref{fig:cs}(a) shows $N(\alpha,0.1n)$ versus $\alpha$ for the unrestricted process, with system size $10^{6}$ and each data point averaged over 100 independent realizations (showing no fluctuations). 
As $\alpha$ first decreases from $\alpha=1$, $N(\alpha,0.1n)$ increases, going from one giant component to two at $\alpha=0.511 \pm 0.003$. Then, once $\alpha<0.11$, $N(\alpha,0.1n)$ decreases.  (Using a less stringent criteria that considers all macroscopic components $C_i n > c n$ where $c>0$ a ``giant", then $N(\alpha,c n)$ actually continues increasing.)

The same reasoning that applied to the original BFW model can be used here to show the stability of the multiple giants. 
Once $k\ge k^*$,  in expectation $P(k,t,n)>\alpha$.  Likewise, once $k\ge k^*$, $\frac{du}{d(k/n)}\sim \mathcal{O}(n^{1/2})$, so $k/n$ increases very slowly and the process frequently samples new links and rejects links that merge any two giants. 
For example, if $\alpha=0.3$, $N(\alpha,0.1n)=3$ with $C_{1}=0.414,C_{2}=0.321,C_{3}=0.265$, so $P(k,t,n)\geq  C_{1}^{2}+C_{2}^{2}+C_{3}^{2}\sim 0.345>\alpha=0.3$ when $k\geq k^{*}(n)$. 
See Fig.~\ref{fig:cs}(b) for details. 

\begin{figure}
\includegraphics[width=0.50\textwidth]{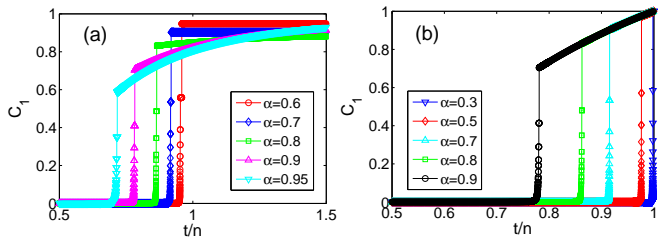}
\caption{
The results of varying $\alpha$ on (a) the unrestricted BFW process in the regime where only one giant component emerges, and (b) the restricted BFW process.  
Both (a) and (b) show increasing delay and larger 
$\Delta C_{max}$ for smaller $\alpha$. 
}
 \label{fig:ds}
 \end{figure}

In Fig.~\ref{fig:ds}(a) we show the typical evolution for the unrestricted BFW process for various values of $\alpha$ in the regime where only one giant component emerges.  We measure the scaling window $\Delta(\gamma,A)$, as discussed earlier, and find that smaller $\alpha$ leads to a more ``explosive" transition in that $A$ is larger and the scaling window shrinks more quickly.  Explicitly, for $\alpha=0.7, 0.8, 0.9$ (and setting $\gamma=1/2$), we find respectively that $A=0.9,0.8,0.7$ and $t_{c}/n \approx 0.915,0.862,0.780$.
This delayed and more explosive nature with smaller $\alpha$  
is intuitive in that the more links are rejected at each stage, the longer one stays in that stage, 
resulting in more components of size $C_i n \sim k$. 

Figure \ref{fig:ds}(b) shows the analogous behavior for the restricted BFW process (where only edges that span components are considered).  
The delayed and more explosive nature of the transition with decreasing $\alpha$ is also observed here. 
We also note that the location of $t_c$ is not affected. For instance, for $\alpha=0.3,0.5,0.7, 0.8,0.9$ we find that $t_{c}/n \approx 0.998,0.976,0.915,0.862,0.780$ for both the restricted and unrestricted processes. 

The behavior of the restricted process can also be explained via Eq.~(\ref{eqn:Pktn}). Here because intra-cluster links are not allowed, the first term on the right-hand side vanishes. If the stage stops at some $k_{0}<n$, then the second term on right side of Eq.~(\ref{eqn:Pktn}) decreases to $0$ which makes $P(k_{0},t,n)<\alpha$ and stage keeps on growing until the two giants merge together. 
The restriction on sampled links does not change the nature of the transition. 
We find that the general BFW model is is strongly discontinuous for all $\alpha \in(0,0.97]$, regardless of whether link-sampling is restricted or unrestricted.

In summary, we have analyzed the critical behavior of the BFW model and find that two stable giant components emerge at the same critical point in a strongly discontinuous transition (Fig.~\ref{fig:bs}(a)). If we restrict the sampled links to only spanning-cluster edges, multiple giants coexist for a few moments until they merge together in a larger discontinuous jump (Fig.~\ref{fig:bs}(a))
 and ultimately only one giant component emerges.  
We further generalize BFW by making the asymptotic fraction of accepted links a parameter $\alpha$, and find that number of stable giant components increases while $\alpha$ decreases. 

The existence of multiple stable macroscopic components is surprising and unanticipated~\cite{SpencerAMS2010}, and has not been previously observed in stochastic percolation. A model of cluster aggregation where largest clusters are occasionally ``frozen" and prevented from growing, does lead to multiple giant components~\cite{Ben-Naim2005}, however the freezing is imposed on the system.   
Simple algorithms that generate multiple giant components may create a new range of applications.  
In addition to providing insight and a potential mechanism for controlling gel sizes during polymerization~\cite{Ben-Naim2005}, 
they may be useful for creating communication networks consisting of multiple large components operating on different frequencies or for analyzing epidemic infections simultaneously arising in distinct, independent groups. 
The unrestricted process allows multiple links between nodes and self-loops.  For finite systems it may be important to understand what happens when only links not yet added to the graph are sampled. In the asymptotic size limit there should be no difference as $O(n)$ edges are added from $O(n^2)$ edges. 

The nature of the transition observed in~\cite{EPScience} was recently analyzed using cluster aggregation models with choice, where a set of candidate edges are simultaneously inspected at each step~\cite{daCostaArxiv}. The mechanism here, in contrast, inspects one edge individually at each time.  As shown in Fig.~\ref{fig:bs}(a) the models in~\cite{daCostaArxiv} (labeled MR (m=2)) and in~\cite{EPScience} (labeled PR) show weakly discontinuous transitions, where $\Delta C_{max}$ decreases with system size.  In contrast both the restricted and unrestricted BFW models are strongly discontinuous, with a jump independent of system size. 
Finally we show the evolution of the component size distribution for the original BFW model in Fig.~\ref{fig:bs}(b) ($n(s)$ is the number of components of size $s$ divided by $n$).  This bi-modal distribution has a large right-hand tail, which deviates from a power law. 

This work was supported in part by the Defense Threat Reduction Agency, Basic Research Award No. HDTRA1-10-1-0088, and the Army Research Laboratory under Cooperative Agreement Number W911NF-09-2-0053.


\begin{thebibliography}{10} 

 \bibitem{T. Bohman and  A. Frieze}
 T. Bohman, A. Frieze, and N. C. Wormald.
 \newblock{\em Random Structures \& Algorithms}, 25(4):432-449, (2004).
 
 \bibitem{StaufferPercBook}
D. Stauffer and A. Aharony.
\newblock {\em Introduction to Percolation Theory} 
(Taylor \& Francis, London, 1994).

\bibitem{Sahimi}
M. Sahimi.
\newblock{\em Applications of Percolation Theory} (Taylor \&
Francis, London, 1994). 

   \bibitem{satorrasPRL}
   R. Pastor-Satorras and A. Vespignani.
   \newblock{\em Phys. Rev. Lett},
   86, 3200 (2001).
    
   \bibitem{moorePRE}
   C. Moore and M. E. J. Newman.
   \newblock{\em Phys. Rev. E},
   61, 5678 (2000).

\bibitem{EPScience}   
D. Achlioptas, R. M. D'Souza, and J. Spencer.
\newblock{\em Science},
323, 1453 (2009). 
   
  \bibitem{ChoPRL09}
 Y. S. Cho, J. S. Kim, J. Park, B. Kahng, and D. Kim.
  \newblock{\em Phys. Rev. Lett.}, 103, 135702 (2009). 
 
 \bibitem{RadFortPRL09}
 F. Radicchi and S. Fortunato.
 \newblock{\em Phys. Rev. Lett.}, 103,168701 (2009). 
 
 \bibitem{ziffPRL09}
 R. M. Ziff. 
 \newblock{\em Phys. Rev. Lett},
 103, 045701 (2009).   
 
 \bibitem{Rziffnew}
 R. M. Ziff, 
 \newblock{\em Phys. Rev. E}, 82, 051105 (2010) 
    
  \bibitem{RaissaMichael}
  R. M. D'Souza, M. Mitzenmacher.
  \newblock{\em Phys. Rev. Lett.}, 104, 195702 (2010)
    
  \bibitem{ChoKahng}
  Y. S. Cho, B. Kahng, and D. Kim.
  \newblock{\em Phys. Rev. E}, 81, 030103(R) (2010)   
  
  \bibitem{Manna}
  S. S. Manna and Arnab Chatterjee.
  \newblock{\em Physica A}, 390, 177-182 (2011).  
  
  \bibitem{Araujo}
  N. A. M. Ara\'{u}jo and H. J. Herrmann.
  \newblock{\em Phys. Rev. Lett.}, 105, 035701 (2010).
  
   
   \bibitem{powderkeg09}  
  E. J. Friedman and A. S. Landsberg. 
  \newblock{\em Phys. Rev. Lett.}, 103:255701, (2009). 
     
   \bibitem{Timme2010}
   J. Nagler, A. Levina, and M. Timme. 
   {\em Nature Phys.} 7 (2011). 
   doi:10.1038/nphys1860.
      
    \bibitem{Makse2009}
   H. D. Rozenfeld, L. K. Gallos, and H. A. Makse.
   \newblock{\em Eur. Phys. J. B}, 75, 305-310 (2010).
         
   \bibitem{SpencerAMS2010}
   J. Spencer. {\em Notices of the {AMS}}, 57 (6) 720-724 (2010).
      
   \bibitem{ER}
   P. Erd\H{o}s and A. R\'{e}nyi.
   {\em Publ. Math. Inst. Hungar. Acad. Sci.}, 5 (17) (1960). 

 	
 \bibitem{Ben-Naim2005}
 E. Ben-Naim, P. L. Krapivsky. 
 {\em J. Phys. A}, 38 (23) L417-L423 (2005). 

 \bibitem{daCostaArxiv}  
  R. A. da Costa, S. N. Dorogovtsev, A. V. Goltsev, J. F. F. Mendes.
    \newblock{\em Phys. Rev. Lett.}, 105:255701, (2010). 
    
\end{thebibliography}
\end{document}